\documentclass[twocolumn]{aastex63}

\newcommand\Lya{Lyman~$\alpha$~}
\newcommand\HI{\ion{H}{1}~}
\newcommand\kms{km~s$^{-1}$}

\acceptjournal{ApJ}
\shorttitle{Correlation between Lyman~$\alpha$ absorbers and HI-galaxies}
\shortauthors{Borthakur}

\begin{document}

\title{How are Lyman~$\alpha$ absorbers in the cosmic web related to gas-rich galaxies?}

\correspondingauthor{Sanchayeeta Borthakur}
\email{sanch@asu.edu}
\author[0000-0002-2724-8298]{Sanchayeeta Borthakur}
\affiliation{School of Earth and Space Exploration, Arizona State University, 781 Terrace Mall, Tempe, AZ 85287, USA}

\begin{abstract}

We present the two-point cross-correlation function between \Lya absorbers and \HI\ galaxies in the nearby Universe ($0.01 \le z \le 0.057$). 
We use absorbers from 21 QSO sightlines from the Survey of the Low-Redshift Intergalactic Medium with HST/COS and the galaxy catalogs from the Arecibo Legacy Fast ALFA survey and the New York University Value-Added Galaxy Catalog.
We find that \Lya absorbers are strongly correlated to \HI\ galaxies at a projected separation of $\le$0.5~Mpc and velocity separation of $\le$50~\kms. 
 \Lya  absorbers are 7.6 times more likely to be found near \HI galaxies compared to a random distribution.  
The correlation decreases as the projected and/or velocity separation increase. We also find the correlation between \Lya\ absorbers and \HI galaxies to be stronger than those observed between \Lya\ absorbers and optically selected galaxies.
There is an enhancement in the number of absorbers at velocity separations of $\le$30~\kms\ from \HI galaxies at distances larger than their viral radius.
Combined with the fact that most of our galaxies are not driving strong outflows, we conclude that the absorbers at low-velocity separations are tracing cooler intergalactic gas around galaxies.
This conclusion is consistent with the predictions from cosmological simulations where faint gas from the intergalactic medium flows into the disks of galaxies leading to galaxy growth.

\end{abstract}


\section{Introduction} \label{sec:intro}

Most of the baryons in the Universe lie in the intergalactic medium (IGM) outside bound gravitational structures like galaxies, groups, and clusters of galaxies \citep[][and references therein]{bregman07, rauch98, mcquinn16}. In our current cosmological model, galaxies are fed by this reservoir either by shock heated gas  \citep[hot mode accretion;][]{white_frenk91, birn03} or via filamentary structures that keep gas cool and prevent it from shock heating \citep[cold mode accretion;][]{keres05, keres09, somerville_dave15}.

Multiple observations suggest that accretion of cool intergalactic gas can support the observed galaxy growth \citep{Sancisi08, fox_dave_17}. For example, the gas distribution around galaxies is known to extend well beyond their viral radii \citep{steidel10,zhu14}, thus suggesting a possible fuel reservior. 
Similarly, the strong correlation between the strength of \Lya absorbers in the outer circumgalactic medium (CGM) of low redshift galaxies and the neutral gas mass fractions of the host galaxies is believed to be a consequence of slow but continuous accretion of gas into galaxies \citep[][]{borthakur15,kauffmann16, kauffmann19, rottgers20}. There is also growing evidence that the neutral hydrogen content of the outer CGM ($\ge$100~kpc) of galaxies at $z\sim2$ is dominated by gas accretion from the IGM \citep{chen_steidel20}.

Since the gas content and star formation activity in galaxies are strongly correlated \citep{catinella10}, the properties of the CGM are expected to vary between star-forming and quiescent galaxies. A comparison of the CGM between blue star-forming and red quiescent galaxies with stellar masses, log~$\rm M_{\star} \sim 9.5 - 11.5$ from the COS-GASS \citep{borthakur15} and COS-Halos \citep{werk14} surveys revealed that the covering fraction of neutral circumgalactic gas as traced by Lyman~$\alpha$ absorbers is $\approx$100\% in blue star-forming galaxies, but lower for red quiescent galaxies \citep{borthakur16}.
Similar differences in \ion{Mg}{2} covering fraction is also seen between red and blue galaxies \citep{anand21}.
The redshift evolution of cool gas content of the CGM is also found to be consistent with the star-formation rate of the host galaxies \citep{lan20}. 
These results indicate that star-forming galaxies are somewhat efficient at acquiring gas, although further investigation is warranted to confirm the physical processes at play.

Outflows from star-forming galaxies are also known to influence the CGM and the intergalactic medium \citep{adelberger05, bouche07, tumlinson11, borthakur13, nielsen16, heckman17}.
In the nearby Universe, most galaxies have low star-formation rate surface density and do not drive strong outflows \citep{heckman_borthakur16, Roberts-Borsani20}. Therefore, at current times outflows from star forming galaxies are expected to enrich only the inner CGM. The outer CGM and gas beyond the viral radius is, therefore, likely to be dominated by cosmic gas flows. Recent semi-analytical modeling by \citet{Afruni20} suggests that outflows can not produce the cool CGM of star-forming galaxies; instead, it is likely a results of gas accretion from the IGM.

While direct detections of gas accretion \citep[e.g.,][]{Sancisi08, bouche13, borthakur19} are rare, statistical  tools such as the two-point cross-correlation function  \citep{davis_peeble83,mo_white96} can provide strong statistical evidence of gas accretion from the IGM into galaxies.
Several studies have characterized the association of galaxies and the neutral gas in the IGM traced as Lyman~$\alpha$ absorber \citep{bouche04,chen05,ryan-weber06, wilman07, lundgren09,chen_mulc09, shone10,tejos14, finn16, liang20} using the two-point cross-correlation function.
Lyman~$\alpha$ absorbers in the spectra of distant Quasi-Stellar Objects (QSOs) are unbiased tracers of neutral gas and are not affected by distance or metallicity. They trace gas in the IGM that lies in the same over-densities where galaxies are but not bound to individual galaxies \citep{cote05, putman06}.

In this study, we investigate the connection between the neutral intergalactic medium and gas-rich galaxies. 
Galaxies that are efficient at accreting gas are expected to be gas-rich systems and hence detectable in \HI 21cm observations. 
Our motivation is to explore if \HI galaxies have access to a potential reservoir of gas that could be feeding them at the current epoch.
To investigate this hypothesis, we evaluate the two-point cross-correlation function between \Lya absorbers and galaxies detected in \HI~21cm. We present our sample and methodology in \S~\ref{sec:sample_methodology} and discuss the results and their implications in \S~\ref{sec:results}. We summarize our findings in \S~\ref{sec:summary} with an outlook towards the future.

The cosmological parameters used in this study are $H_0 =70~{\rm km~s}^{-1}~{\rm Mpc}^{-1}$ (in between the two recent measurements of $\rm 73.24 \pm 1.74 ~ km~s^{-1}~Mpc^{-1}$ \citep{riess16} and $\rm 67.6_{-0.6}^{+0.7}~{\rm km~s}^{-1}~{\rm Mpc}^{-1}$ \citep{grieb16}), $\Omega_m = 0.3$, and $\Omega_{\Lambda} = 0.7$. A 5--10\% variation in the value of the Hubble constant or the cosmology would not affect the overall conclusions of this paper.

\section{Sample and Methodology \label{sec:sample_methodology}}

\subsection{The \Lya Absorber Catalog}

We have constructed our absorber catalog by selecting \Lya absorbers from the Survey of the Low-Redshift Intergalactic Medium with HST/COS\footnote{\url{https://archive.stsci.edu/prepds/igm/}} \citep{danforth16}. We chose a sub-sample of 21 QSO sightlines that lie within the sky coverage of the Arecibo Legacy Fast ALFA survey \citep[ALFALFA;][]{alfalfa}  and the Sloan Digital Sky Survey \citep[SDSS,][]{york00}. We also chose a redshift range of $0.01 \le z \le 0.057$ for the sample based on the following considerations. The lower end was constrained by our ability to detect \Lya absorbers in the QSO spectra near the damping wings of the galactic \Lya profile ($\lambda \ge  1228 \rm \AA$). The higher end of the redshift range was determined by the velocity coverage of the ALFALFA survey ($v\le 18000$~km~s$^{-1}$).

This yielded a sample of 124 \Lya  absorbers, with 80\% of them at column densities lower than log(N(HI))$<$14.0.
These low column density absorbers are believed to be tracing the gas in the IGM \citep{chen_mulc09} and hence are likely to trace cosmic gas flows.
The \citeauthor{danforth16} catalog selected QSO sightlines that have spectra with a signal-to-noise ratio S/N $>$ 15 per resolution element of $\sim$ 17~\kms. This results in good coverage of the weak absorbers. On the other hand, the catalog is biased towards bright background sources. This may result in selecting QSO that may have lower numbers of high column density absorbers. Those absorbers are known to be closely correlated to galaxies \citep[e.g.][]{chen05}, potentially tracing the CGM. Therefore, our cross-correlation function may be weaker than that measured from a sample of flux independent QSO sightlines.

\newpage

\subsection{The Galaxy Catalog}

We use two galaxy catalogs in our study. The first is the ALFALFA extragalactic HI source catalog data release of March 2018\footnote{http://egg.astro.cornell.edu/alfalfa/data/} \citep{haynes18}, which detected galaxies in the nearby Universe via the \HI 21~cm hyperfine transition. As a result, galaxies seen in this survey contain a significant amount of neutral interstellar medium. The ALFALFA survey is particularly sensitive to dwarf or gas-rich galaxies that may or may not have a significant stellar component. In our analysis, we make use of all the galaxies within 10~Mpc of the 21 QSO sightlines in the redshift range $0.007 \le z \le 0.06$ bracketing the absorber redshift range by $\pm$1000~\kms.
These criteria yield a total of $\sim$ 8000 galaxies with \HI masses ranging from $\rm log~(M_{HI}/M_{\odot})= 7.8 - 10.9$. 

We also use the New York University Value-Added Galaxy Catalogue \citep[NYU-VAGC,][]{blanton05} derived from the SDSS spectroscopic data release 7 \citep{SDSS_DR7}. Being an optically derived catalog, it is complementary to the ALFALFA catalog. While the ALFALFA survey is biased towards gas-rich galaxies, the NYU-VAGC is biased towards optically brighter galaxies. Comparing the cross-correlation function between these two galaxy catalogs can provide crucial insights into the connection between the observed properties of galaxies and the IGM that surround them.

\begin{figure*}[]
\center 
\includegraphics[trim = 0mm 0mm 0mm 0mm, clip,angle=-0, width=7.15in]{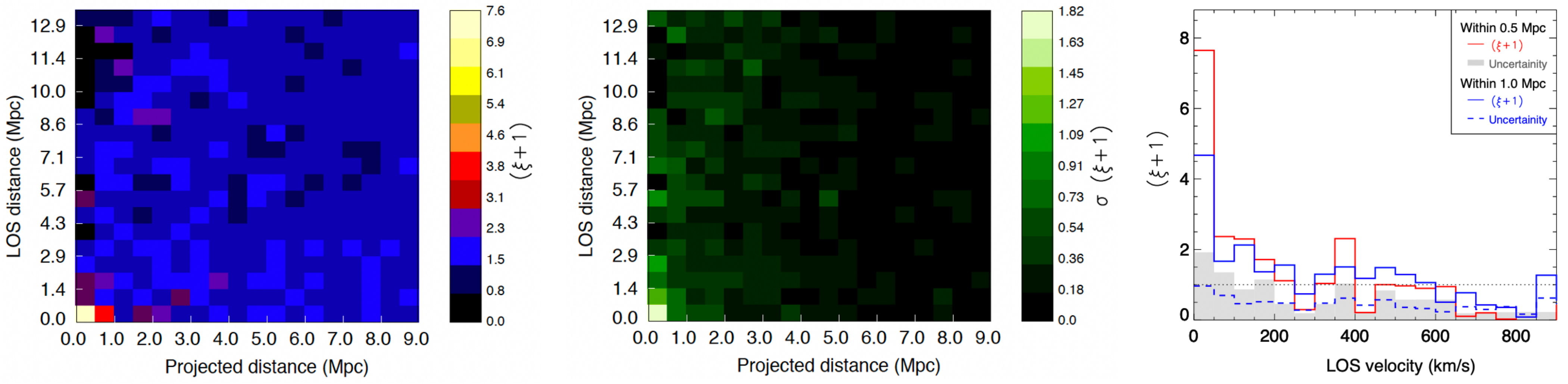}
\caption{The left panel shows the two-dimensional cross-correlation function between \Lya absorbers and ALFALFA galaxies. The peak of the correlation function, $\xi+1=$7.6, was observed at the smallest bin, i.e., within 0.5 and 0.7~Mpc from galaxies.  The line-of-sight (LOS) distance was calculated assuming a pure Hubble flow and H$_0= $ 70~\kms\ from a bin size of 50~\kms. The middle panel shows the uncertainties in the cross-correlation function estimated using bootstrap sampling. The right panel shows slices of the cross-correlation function in terms of line-of-sight velocity for galaxies within 0.5~Mpc and 1~Mpc in red and blue, respectively. The uncertainties are shown as the gray shaded area and dashed blue line corresponding to the red (0.5~Mpc) and blue (1~Mpc) slices. }
 \label{CC-ALFALFA} 
\end{figure*}

\subsection{The Absorber Galaxy Cross-correlation Function}

The cross-correlation function, a tool to assess the association between the galaxy and the absorber populations, is defined as
 \begin{equation} 
dP=ndV [1+\xi_{AG}({\bf r_2-r_1})]
\end{equation} 
where, dP is the conditional probability of finding a galaxy at position {\bf r$_2$} within volume dV, given there is an absorber at {\bf r$_1$}, $n$ is the unconditional galaxy density, and $\xi_{AG}$ is the cross-correlation function. We evaluate the cross-correlation function using the estimator developed by \citet{Landy_Szalay93} given by
\begin{equation} 
\xi_{AG}(r_{pd}, r_{los})=\frac{A_dG_d - A_dG_r - A_rG_d }{A_rG_r } +1
\end{equation} 
where, $r_{pd}$ is the projected distance, $r_{los}$ is the line of sight distance, AG refers to number of absorber galaxy pairs, and the subscripts {\em d} and {\em r} refer to data and random catalogs respectively. The $ r_{los}$ was derived from converting the redshift into distance assuming a Hubble flow and neglecting contributions from peculiar velocities.

The AG pairs were estimated by counting the number of galaxy-absorber pairs within each projected distance and line-of-sight distance bin. The random galaxy catalog was generated by assigning each galaxy a redshift that was randomly sampled from the distribution of redshifts for the galaxy catalog. This is similar to procedure used by \citet{adelberger03}, \citet{tejos14}, and \citet{finn16}. Care was taken to draw redshifts from a distribution with same selection criteria (luminosity, color, mass, etc.) as the true galaxy catalog. 
The  random catalogs for the absorbers were created by distributing the absorbers randomly within the redshift range of the true data except in for the ranges that are inaccessible for detecting redshifted \Lya in the QSO spectra due to the presence of saturated absorption lines from the Milky Way. 
The random pairs ($A_rG_r,  A_dG_r,  A_rG_d$) were generated by creating 100 iterations of the random catalogs -- both galaxy and absorber catalogs wherever applicable -- and then normalizing them to to get the final random pair counts.

The cross-correlation function was evaluated at a velocity resolution, $\delta \pi$= 50~\kms ~corresponding to 0.7~Mpc out to 13.6~Mpc. 
This choice was motivated by the resolution of the SDSS spectra that were used for galaxy redshift identification for the NYU-VAGC. 
The ALFALFA survey has a higher spectral resolution, so finer binning was possible; however, the galaxy counts in ALFALFA survey are much lower than SDSS per unit volume resulting in a low signal-to-noise (S/N) cross-correlation function. Therefore, for easy comparison with SDSS and for optimal S/N, we chose a line-of-sight (LOS) binning resolution of 50~\kms.
To roughly match the line-of-sight bin spacing, we chose a bin size of $\delta \sigma =$ 0.5~Mpc for the projected separation axis.

We estimated the uncertainties in the cross-correlation function using block bootstrap, where spatial blocks of data are randomly sampled \citep{Carlstein98, holmes_reinert04, Loh08, Loh_stein08, feigelson_babu_2012,Baddeley15}. 
Bootstrap sampling is commonly used to estimate the bias and standard error of a statistic when it is generated from a random sample.
The bootstrap sampling was done with 1000 samples of the correlation function each derived from 21 sightline randomly sampled (with replacement) from the original 21 sightlines. 
For further details on the bootstrap sampling procedure and its appropriateness for estimating uncertainties in the cross-correlation function, we refer the reader to the discussion by \citet[][and references therein]{finn16}.

\newpage

\section{Results and Their Implications \label{sec:results}}

\subsection{The strength of the cross-correlation function and origin of the correlation }

The cross-correlation function ($\xi +$1) for \Lya absorbers and ALFALFA galaxies is presented in the left panel of Figure~\ref{CC-ALFALFA}. The cross-correlation strength peaked at 7.6$\pm$1.9 at the nearest bin covering a projected separation of $\le$0.5~Mpc and velocity separation of $\le$50~\kms, corresponding to $\le$0.7~Mpc. We detected 11 true galaxy-absorber pairs as compared to 1.2--1.5 random pairs ( for $ A_rG_r, A_dG_r$, and $A_rG_d$)  expected from a random distribution.
The true pairs were generated by 8 absorbers with column densities ranging from log[N(HI)] = 13.13 $-$ 15.61. The strongest among them with log~N(HI) =15.61 had three galaxies within 0.5~Mpc and 50~\kms, whereas the second strongest absorber with log N(HI) =14.16 had two galaxies in the same range of projected and velocity separation. The remaining 6 absorbers were matched to one galaxy each.

The strength of correlation dropped rapidly with increasing projected and line-of-sight separation. In bins adjacent to the peak, the strength was merely 20--33\% of the peak.
The drop continued with projected separation and line-of-sight distance to reach levels close to $\xi +$1 = 1 beyond 2~Mpc, implying a random distribution of absorbers with respect to galaxies. The uncertainties associated with the cross-correlation function are presented in the middle panel of Figure~\ref{CC-ALFALFA}. A velocity slice of the cross-correlation function is shown in the right panel of Figure~\ref{CC-ALFALFA} for the nearest projected separation bin of 0.5~Mpc and the first two bins combined, i.e., out to 1~Mpc.
The plot illustrates the drop in the strength of correlation as a function of line-of-sight velocity. We do not find a significant signal beyond 200~\kms, although there is a hint of mild finger-of-god effect.

\begin{figure*}[t]
\center 
\includegraphics[trim = 0mm 0mm 0mm 0mm, clip,angle=-0, width=7.15in]{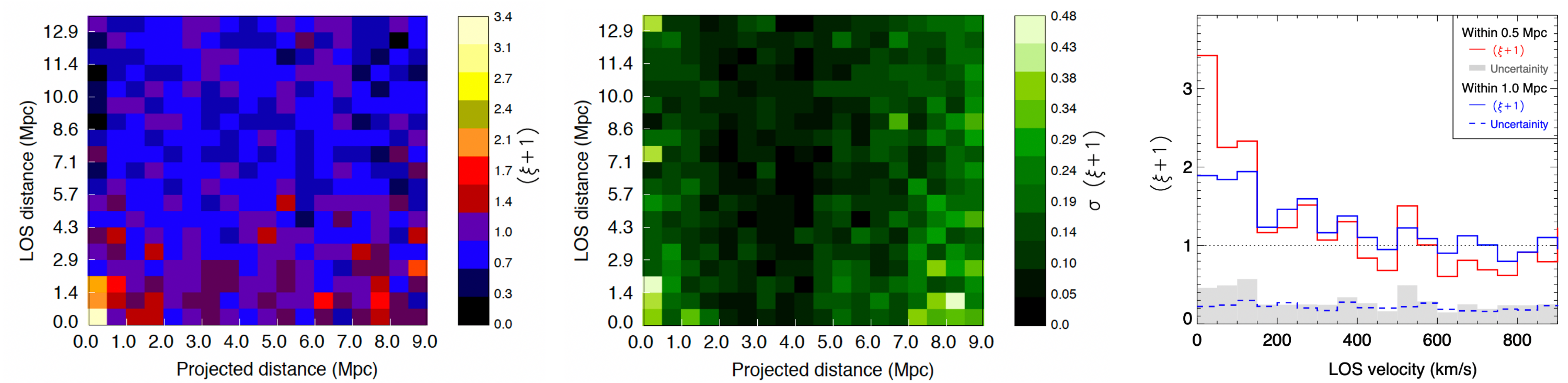}
\caption{The left panel shows the two-dimensional cross-correlation function between \Lya absorbers and galaxies from the SDSS-derived NYU-VAGC. The peak of the correlation function, $\xi+1=$3.4, was observed at the smallest bin, i.e., within 0.5 and 0.7~Mpc from galaxies. The line-of-sight (LOS) distance was calculated assuming a pure Hubble flow and H$_0= $ 70~\kms\ from a bin size of 50~\kms. The middle panel shows the uncertainties in the cross-correlation function estimated using bootstrap sampling. The right panel shows slices of the cross-correlation function in terms of line-of-sight velocity for galaxies within 0.5~Mpc and 1~Mpc in red and blue, respectively. The uncertainties are shown as the gray shaded area and dashed blue line corresponding to the red (0.5~Mpc) and blue (1~Mpc) slices. }
 \label{CC-NYU} 
\end{figure*}

A similar analysis with the SDSS spectroscopically confirmed galaxies from the same QSO fields using the NYU-VAGC is presented in Figure~\ref{CC-NYU}. Qualitatively, the results were similar to the runs for the ALFALFA galaxies. 
The peak value of the correlation function was $\xi +$1=3.4, half of the peak observed in \HI\ galaxies.
The peak strength was observed at the smallest projected separation and los distance. The strength dropped rapidly as a function of projected separation and line-of-sight velocity.
Beyond 150~\kms, the cross-correlation strength diminished to that of a random distribution.
The uncertainties quoted in redshift measurements of the main SDSS spectroscopic catalog\footnote{https://classic.sdss.org/dr7/} is 30~\kms \citep{Strauss02}. For  galaxies in our redshift coverage ($0.01 \le z \le 0.057$), we measured the median uncertainty in redshift to be ${\tilde{\delta z}}= 0.000117$ corresponding to a velocity of 35~\kms. Therefore, the spread of 150~\kms\ is likely to be an intrinsic property of the distribution.
On the other hand, there is a substantial difference between the number density of galaxies in the NYU-VAGC and the ALFALFA catalog. So it is likely that the possible clustering of gas-poor galaxies with gas-rich galaxies would show up in the run with the optical sample.
The difference in the galaxy number density between NYU-VAGC and ALFALFA also impacts the uncertainties in the cross-correlation signal -- the uncertainties in the NYU-VAGC runs are about a third of the uncertainties in the ALFALFA run.
Consequently, the correlation between optical galaxies and \Lya\ absorber is detected at a higher signal-to-noise ratio than the correlation between \HI\ galaxies and \Lya\ absorber, although the absolute strength of the latter is higher.

We also estimated the cross-correlation function for sub-catalogs derived from the NYU-VAGC to investigate its dependance on the stellar mass and star-formation of the galaxies. The three sub-catalogs were: 
\begin{enumerate}

\vspace{-0.30cm}

\item Galaxies with stellar mass M$_{\star} \ge \rm 10^{7.8}~M_{\odot}$ corresponding to the \HI mass detection limit of the ALFALFA catalog;

\vspace{-0.3cm}

\item Galaxies with stellar mass, M$_{\star} \ge \rm 10^{7.8}~M_{\odot}$ that are labeled as ``emission-line'' dominated in the SDSS spectral classification; and 

\vspace{-0.25cm}

\item Galaxies with stellar mass, M$_{\star} \ge \rm 10^{7.8}~M_{\odot}$ that are defined as ``blue" galaxies based on their position in the galaxy color bimodality diagram, with colors derived from broad-band photometry.

\end{enumerate}

\vspace{-0.1cm}

The galaxy stellar mass and broad-band K-corrected photometry were adopted from the NYU-VAGC. About 40\% of the galaxies in the catalog have M$_{\star} \ge \rm 10^{7.8}~M_{\odot}$. We adopted the galaxy-color definition by \citet{vandenbosch08}, which was derived from the NYU-VAGC. 
Following their procedure, the galaxies were divided into two categories - blue (A$\le$0.66) or red(A$>$0.66) - based the value of the parameter $ A = (g-r)-0.1[\rm log~M_{\star} - 10]$.

\begin{figure}[]
\center 
\vspace{.3cm}
\includegraphics[trim = 0mm 0mm 0mm 0mm, clip,angle=-0, width=2.73in]{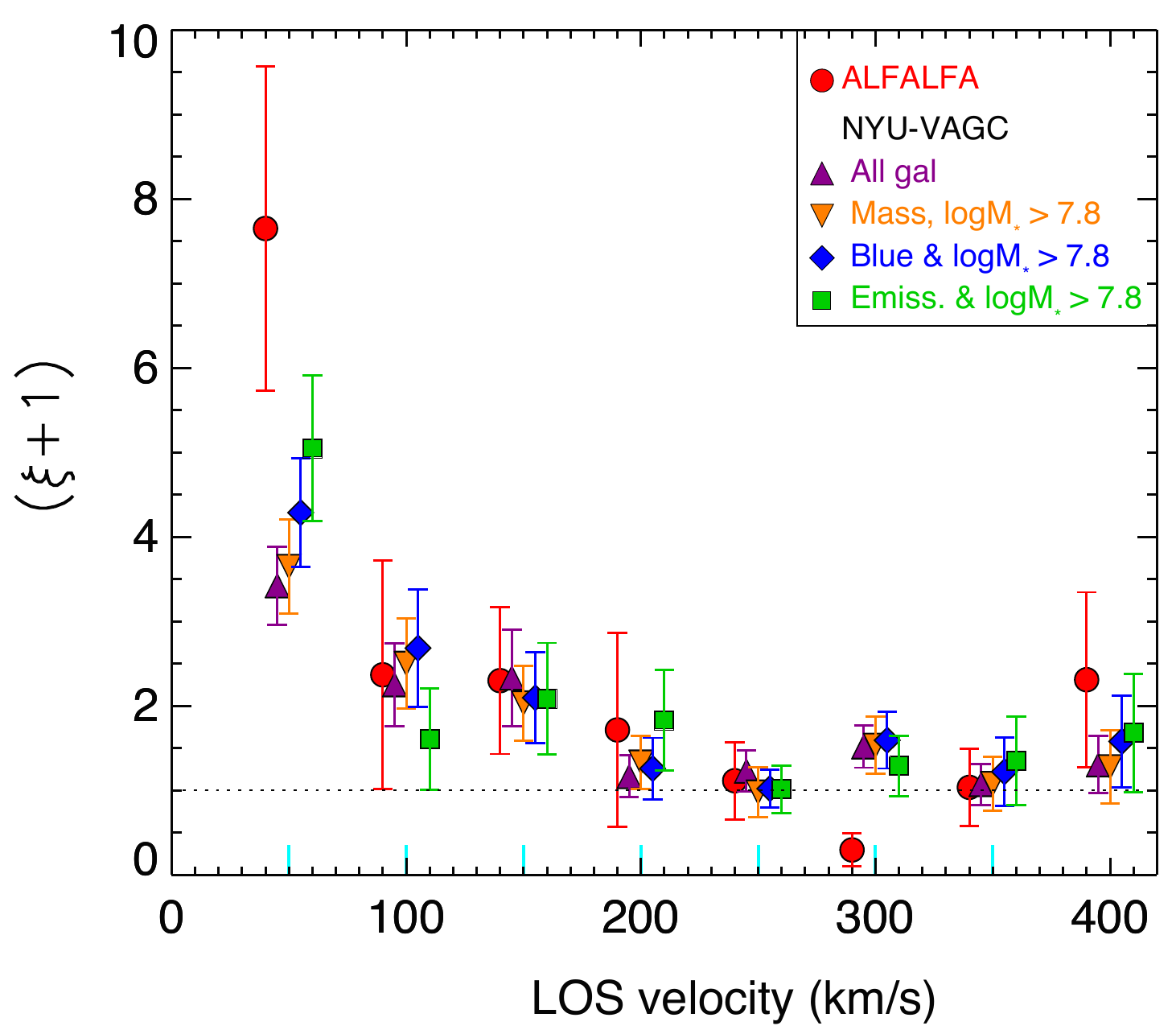}
\caption{Comparison of the strength of cross-correlation function for runs with the ALFALFA catalog and sub-catalogs derived from the NYU-VAGC. The points show the variation in the strength of the cross-correlation function as a function of line-of-sight (LOS) velocity for the nearest projected separation bin of $\le$0.5~Mpc. The peak strength of correlation was observed at LOS velocity $\le$50~\kms\ for all the runs with varying strengths. The strongest correlation was seen for ALFALFA galaxies followed by emission-line galaxies for the same set of \Lya\ absorbers. The dotted line at $\xi+1=1$ indicates no correlation. }
\label{CC-comp} 
\end{figure}

The cross-correlation function for all the three sub-catalogs show similar overall trend. In all cases, the peaks were observed at the bin nearest to the sightline both in terms of projected distance and line of sight separation. 
Figure~\ref{CC-comp} shows a comparison of the strength of the cross-correlation function for all the runs as a function of line-of-sight velocity for a projected separation of 0.5~Mpc. The relatively large uncertainties in the strength of the cross-correlation function made it difficult to distinguish significant variations between the runs.

The strength of the correlation was found to be stronger between \Lya absorbers and \HI galaxies than optically selected galaxies. 
Among the optically selected galaxies, the emission-line galaxies showed the strongest correlation to \Lya absorbers. A similar result was reported by \citet{chen05}. 
The trends in the cross-correlation strength seen among the various samples are consistent with our understanding of cosmological gas accretion into galaxies. The gas from the IGM flows into the halos of galaxies, feeds their \HI\ disks, which leads to star formation.
The time lag between these steps is likely to manifest as a reduction in correlation strength between \Lya\ absorbers and galaxy populations selected based on gas mass or star-formation indicators.
 In this picture, the warm gas in the IGM would correlate the most with \HI\ galaxies, followed by emission-line galaxies with active \ion{H}{2} regions indicating the presence of young ($<$10~Myr) O stars, and then with blue galaxies that have gone through a star-formation phase in the last few Gyrs.

In conclusion, we found the  \HI\ galaxies to correlate strongly with the absorbers than any other sample considered in this study. This would hint that gas-rich galaxies are surrounded by an IGM richer in \Lya\ absorbing clouds.  
We infer that these clouds are either stationary or flowing toward the galaxies and do not trace outflows. 
First, the velocity range of the strong correlation signal is extremely small, i.e., $\le 50$~\kms, which is less than the escape velocity for most galaxies. 
Second, the star-formation rates of low-redshift galaxies are not enough to drive strong outflows that can escape the gravitational potential of the galaxy \citep{Heckman05}. 
These two facts make it difficult for the galactic outflows scenario to generate a population of absorbers with low-velocity offsets out to 1~Mpc.

\subsubsection{Individual absorber-galaxy pairs}

Among the 124 \Lya absorbers spread over a redshift range corresponding to 14,000~\kms, we found 8 absorbers to have at least one \HI detected galaxy in the ALFALFA catalog within 0.5~Mpc from their positions with velocity separation of $\le$50~\kms. 
The number of absorbers that have at least one galaxy within a Mpc and the same velocity separation increased to 14. While the sky coverage (area) between the two regions of radii 0.5 and 1~Mpc is 1:4, the number of absorbers with a neighboring \HI galaxy was 1:1.75. This indicated that \Lya\ absorbers are clustered around \HI\ galaxies and not randomly distributed.

We also found that absorbers tend to have velocities comparable to the nearest galaxies. For example, we detected 17 absorbers with an \HI\ galaxy within 0.5~Mpc and 500~\kms, 8 of which had velocity separations of less than 30~\kms. In other words, 47\% of the absorbers cluster in 6\% of the velocity space around the \HI\ galaxies.
The propensity of absorbers to be close to the galaxy systemic velocity is seen out to 1~Mpc, where more than 30\% of the absorbers have velocity offsets of $\le$ 30~\kms\ (out of total of 38 absorbers with a galaxy within 1~Mpc and $\delta v \le$ 500~\kms). 
To put that in context, the number of absorber-galaxy pairs with $\le$10~Mpc in physical separation and $\le$500~\kms\ in velocity separation is 112. Only 15 of those show a velocity offset of $\le$ 30~\kms, corresponding to 13\%. Therefore, there is an enhancement in the number of absorbers with low-velocity separation at low projected distances from \HI\ galaxies. Among the absorbers with velocity offsets, $|\delta v| \le 30$ \kms, 8 were at projected separation of $\le$0.5~Mpc, 4 were at projected distance between 0.5--1~Mpc, and only 3 were at projected distance between 1--10~Mpc. This distribution is not consistent with a spatially uniform distribution of absorbers which should scale with the area on the sky and show a ratio of 1:3:396 instead of the observed 8:4:3.

  \begin{figure}[]
\center 
\vspace{.3cm}
\includegraphics[trim = 0mm 0mm 0mm 0mm, clip,angle=-0, width=2.73in]{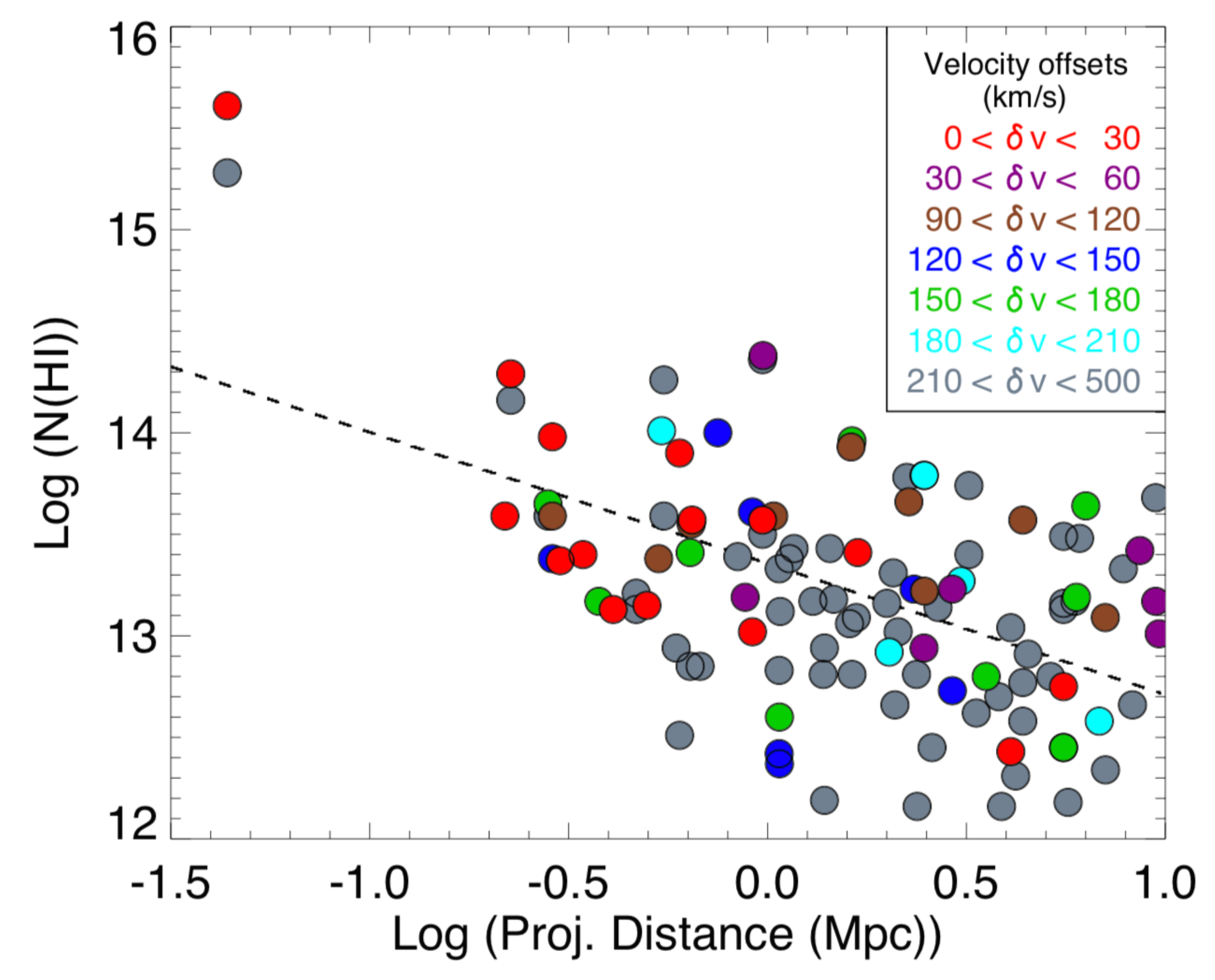}
\caption{The projected separation between the sightline and the closest galaxy for \Lya absorbers in our sample. Uncertainties associated with the column densities are of the order of 0.1 dex, which is smaller than the size of the points and is not plotted here. The dashed line shows the fit to the data and has a slope of $-$0.65 ($\pm$0.1) and an intercept of 13.36 ($\pm$0.05). The velocity separation between the absorber from the nearest galaxy is color-coded. Each color represents a velocity spread of $\pm$30\kms\ except for gray that covers a range of $\pm$290\kms. This explains the large fraction of gray points in the plot. We find a propensity of absorbers within 1~Mpc of the sightlines to have small velocity offset from the galaxy systemic. 10\% of the absorbers (12/124) did not have an \HI galaxy within a separation of 10~Mpc and 500~\kms.  }
\label{CC-abs} 
\end{figure}

The distribution of the projected separation to the nearest galaxies as a function of absorber strength is shown in Figure~\ref{CC-abs}. The points are color-coded to reflect the velocity offset between the absorbers and their nearest galaxy. 
A linear fit to the data is shown as the dotted line with a slope of $-$0.65$\pm$0.1 and an intercept of 13.36$\pm$0.05.
The Spearman's rank correlation coefficient, $\rho = -0.42$ with a significance of 99.99\%.
The plot includes all the absorbers in the sample that have a galaxy within 10~Mpc. 
We see a relatively larger fraction of points within 1~Mpc that have the smallest velocity separation i.e. $\delta v \le $30 \kms. We note that these points primarily lies between projected distances of 200~kpc and 1~Mpc, which would correspond to the outer CGM or beyond the viral radius of a Milky-Way like galaxy with stellar mass $\rm \sim 10^{10} M_{\odot}$. 
This would imply that the measured velocity of the absorbers is dominated by the Hubble flow and not peculiar velocities that generally trace strong ``flows" of gas into massive halos or out of the intermediate-mass galaxy-sized halos. Assuming equipartition of velocity, we conclude that on an average, the bulk of the gas in the cosmic web is moving at velocities $\le$ 52~\kms, which resulted in most of the absorbers being observed within a line-of-sight velocity separation of $\le$ 30~\kms.

\subsection{Comparison to previous studies} \label{prev_studies} 

Several studies have explored the cross-correlation function between \Lya\ absorbers and galaxies, and each of them had its own unique advantages and biases. This makes a direct comparison of the strength of the correlation between different studies difficult to interpret. For example, our UV absorption-line data were obtained with the Cosmic Origins Spectrograph (COS), which has higher sensitivity than the Space Telescope Imaging Spectrograph (STIS) that was used by some of the earlier studies. 
On the other hand, the spectral resolution of COS is half that of STIS, which makes COS data susceptible to blending. Similarly, the galaxy catalogs used among the studies also vary not only in depth but also in the tracer used to detect the galaxies. Considering these differences, the closest study to our is that by \citet[][RW06 hereafter]{ryan-weber06} where \Lya absorbers from STIS day were correlated with galaxies from the \HI Parkes all-sky Survey (HIPASS). HIPASS is a blind \HI survey just like ALFALFA and, in principle, should trace a similar population of galaxies as the ALFALFA survey. 
The difference between our study and that of RW06 is that the ALFALFA survey is about eight times more sensitive and has an angular resolution (FWHM) four times better than HIPASS. Therefore, the galaxy detection rate is an order of magnitude better than HIPASS \citep{martin10}, which makes it particularly suitable for evaluating the cross-correlation function.

Other related studies with different galaxy catalogs and QSO sightline selections criteria were those by \citet[][C05 hereafter]{chen05} and \citet{williger06}. C05 studied the cross-correlation of \Lya\ absorber and optically detected galaxies towards a single sightline (PKS~0405$-$123). They covered a much larger redshift space of $z=0-0.5$, and their galaxy catalog was 90\% complete to an R-band magnitude of 20. \citeauthor{williger06} also studied the same field with a slightly less homogenous galaxy catalog.

\begin{figure}[]
\center 
\vspace{.3cm}
\includegraphics[trim = 0mm 0mm 0mm 0mm, clip,angle=-0, width=3.0in]{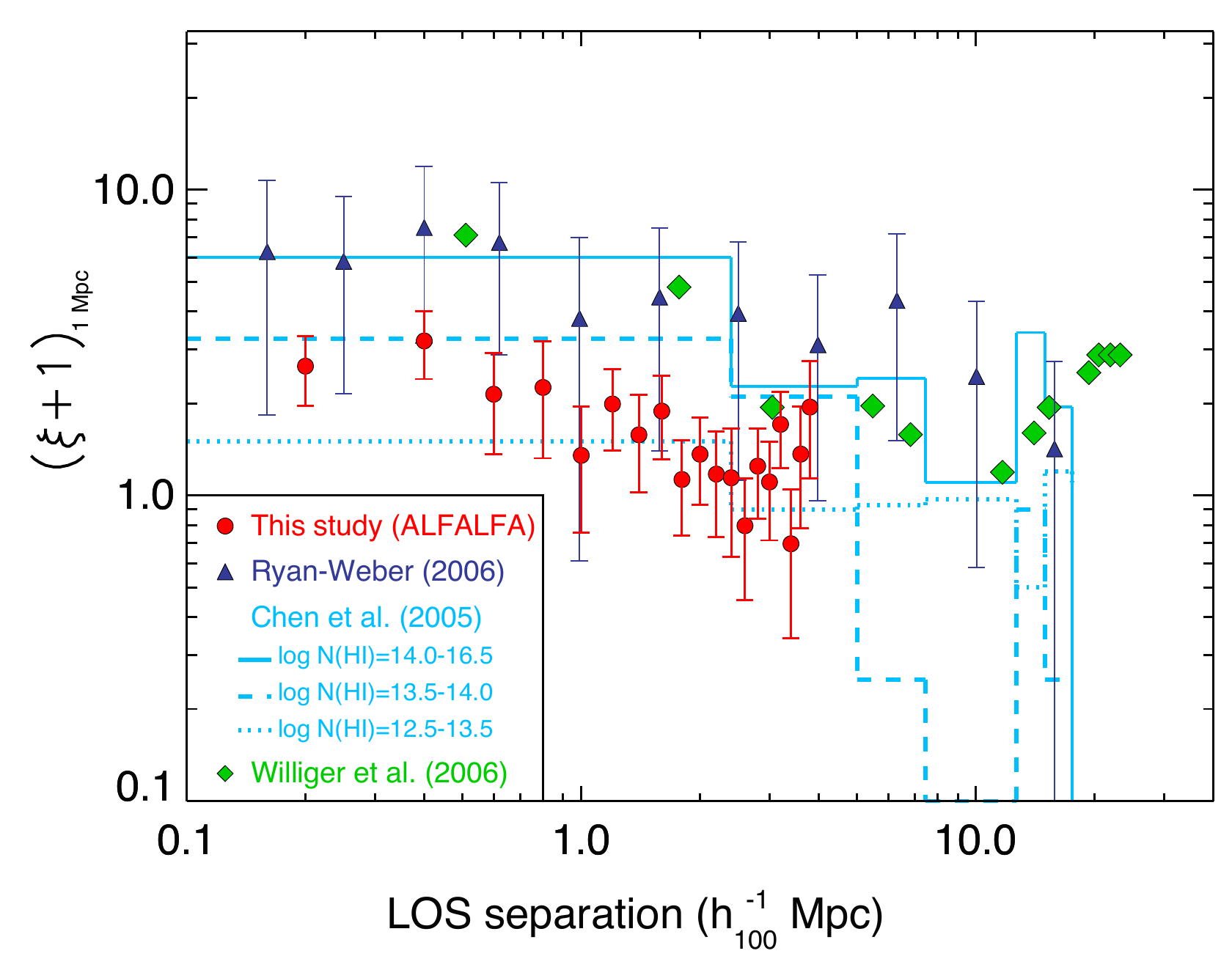}
\caption{The observed cross-correlation strength with a projected separation of 1~$h_{100}^{-1}$~Mpc as a function of line-of-sight separation. The four studies presented in the plot are identified by their symbol shape and color. The red filled circles show our estimation using the full ALFALFA catalog (similar to shown in Figure~\ref{CC-ALFALFA})  a projected separation of 1~$h_{100}^{-1}$~Mpc and assuming a Hubble constant of 100~\kms~Mpc$^{-1}$. This was done to match the binning scheme of the studies by RW06, C05, and \citet{williger06} so that meaningful comparison can be made. cross-correlation measurements from \citeauthor{chen05} for \Lya\ absorbers with column densities between log N(HI) of 14.0--16.5, 13.5--14.0, and 12.5--13.5 are shown as solid, dashed, and dotted blue lines, respectively. The other two studies do not evaluate separate cross-correlation functions for different subsamples of absorber column densities. The average uncertainties in \citeauthor{chen05} and \citeauthor{williger06} measurements are $\pm$1.5 to $\pm$2. We found that our measurements are consistent with those found by RW06 and C05. Beyond 2~$h_{100}^{-1}$~Mpc, all the studies showed slight divergence from each other, most likely dominated by uncertainties in the measurements. The strength of the correlation, $\xi +1 $ approaches 1 indicating no correlation at line-of-sight sedation of  5~$h_{100}^{-1}$~Mpc. }
 \label{other_studies} 
\end{figure}

To compare our results to these studies, we re-evaluated the cross-correlation function with a different bin size in projected separation of 1~$h_{100}^{-1}$~Mpc. 
Figure~\ref{other_studies} shows the line-of-sight variation of the cross-correlation strength for absorber-galaxy pairs within 1~$h_{100}^{-1}$~Mpc of projected separation. Our cross-correlation function for \HI\ selected  ALFALFA galaxies is shown as filled red circles along with the measurements by RW06 (dark blue triangles), C05 (light blue lines), and \citeauthor{williger06} (green diamonds). Qualitatively, our results broadly match those by C05, although the galaxy catalog used in their study is optically selected.
At low LOS separations, our values match C05's run with N(HI) $=13.5-14.0$ whereas at higher LOS separations they match C05's run with N(HI) $=12.5-13.5$. This is not surprising as most of our absorbers have log~N(HI) $<$14.0, and hence our cross-correlation function is dominated by absorbers with column densities, log~N(HI) $\sim 13$.
Our measurements are also consistent with RW06 and fall within their uncertainties for most LOS bins.

One of the crucial aspects to recognize here is that our cross-correlation function gets diluted when integrating over a larger projected separation of $\approx$1~$h_{100}^{-1}$~Mpc. This can be seen by comparing the peak of this run to that shown in Figure~\ref{CC-ALFALFA}. We believe that the match between the various studies (Figure~\ref{other_studies}) with different data quality and biases may be due to `smoothing' of the cross-correlation function.
The smoothing process is potentially hiding interesting physics between different samples as suggested by the fact that we do find variations in the cross-correlation function within the two smallest bins among our ALFALFA and NYUVAGC runs. 
A solution would be to run an analysis on a much larger sample that can provide resolution without compromising on the signal-to-noise.
Future galaxy surveys such as the ASKAP \HI All-Sky Survey \citep{Koribalski20}, the Commensal Radio Astronomy FasT Survey \citep[CRAFTS,][]{Zhang19}, and other large surveys in the \HI\ 21cm hyperfine transitions with more extensive sky coverage would enable further exploration of the connection between gas-rich galaxies and the IGM.

\section{Summary \label{sec:summary}}

We evaluated the correlation of \Lya\ absorbers with \HI-detected galaxies from the ALFALFA survey. Our absorber sample came from 21 QSO sightlines from the Survey of the Low- Redshift Intergalactic Medium with HST/COS that fall within the sky coverage of the ALFALFA survey and the SDSS. These two galaxy redshift surveys act as our galaxy catalogs for evaluating the cross-correlation function. We summarize our findings below:

\vspace{-0.25cm}

\begin{enumerate}

\item The \HI$-$detected galaxies are strongly correlated to \Lya absorber. The \Lya absorbers are 7.6 times more likely to be found near an \HI- galaxy than at a random position in the Universe. The correlation strength peaked at a physical separation of $\le$0.5~Mpc and a line-of-sight velocity separation of $\le$50~\kms. The strength of correlation dropped rapidly as the separation in either dimension increases.

\vspace{-0.25cm}

\item The \HI galaxies showed a stronger correlation with \Lya absorber than the optically selected galaxies including subsets like emission-line galaxies and blue galaxies. In all these cases, the peak of correlation between absorbers and galaxies, although at different strengths, was seen at the smallest bin that is closest to the absorber both in terms of physical separation and line-of-sight velocity.

\vspace{-0.25cm}

\item The \Lya absorbers were found to cluster around \HI\ galaxies at small velocity separation of $\le$30~\kms. This contributed to the signal at the smallest bin of the cross-correlation function. Assuming equipartition of velocity, this would imply a true motion of the absorbers $\le$ 52~\kms. The propensity of absorbers to be close to the galaxy systemic velocity is strongest for \HI galaxies than optically selected galaxies.

\vspace{-0.25cm}

\item In the absence of strong evidence of outflows generated by low-redshift galaxies that exceed escape velocities, the small velocity offsets between \Lya\ absorbers and galaxies out to 1~Mpc suggests that \HI galaxies are surrounded by neutral gas-rich IGM. Since \HI galaxies showed stronger cross-correlation strengths compared to optically selected galaxies, we conclude that \HI galaxies are in environments richer in neutral gas than optical galaxies. This conclusion hints toward the possible flow of gas from the IGM into the \HI galaxies resulting in these galaxies having a significant neutral ISM to be detected in \HI surveys.

\vspace{-0.25cm}

\item Our estimate of the cross-correlation function between \HI galaxies and \Lya absorber at a lower resolution of 1~$h_{100}^{-1}$~Mpc is consistent with previous results by \citet{ryan-weber06}, and \citet{chen05}. In terms of the galaxy catalog, our study is the closest to the study by \citeauthor{ryan-weber06} and our estimate of the cross-correlation function is within their uncertainties. The improvement in our study is that we are able to reduce the uncertainties significantly and, consequently, a derive a higher signal-to-noise measurement.

\end{enumerate}

In conclusion, we find that the stronger correlation between \Lya absorbers in the IGM and the \HI galaxies confirm that the IGM surrounding \HI\-rich galaxies is richer in neutral hydrogen than that surrounds optically selected galaxies. The small line-of-sight velocity separation between absorbers and galaxies indicate that we are witnessing primarily Hubble flow and not strong peculiar velocities. These observations may hint towards gradual accretion of gas from the IGM onto the \HI galaxies at current times, although the data do not provide direct evidence of gas flow.
In the future, we hope that
Future \HI surveys with additional sky coverages will enable further exploration of the connection between the absorbers and \HI galaxies and lead to a better understanding of cosmological gas accretion into galaxies.

\acknowledgments
We thank the referee and the AAS statistician for their constructive comments.
We thank Charles Danforth, Brian Keeney, Hsiao-Wen Chen,  Daniel McIntosh, Xavier Prochaska, Nicolas Tejos, Todd Tripp, and members of the STARs Lab at ASU --  Chris Dupuis, Tyler McCabe, Mansi Padave, Jackie Monkiewicz, Hansung Gim, and Ed Buie II  -- for various discussions over the years that lead to this work. 
SB also thanks Anika B. Srivastava for her unique suggestions to improve the presentation of the figures. 
SB also acknowledges the indigenous peoples, including the Akimel O'odham (Pima) and Pee Posh (Maricopa) Indian Communities, whose care and keeping of the land has enabled her to be at ASU's Tempe campus in the Salt River Valley, where this work was conducted.

\bibliographystyle{apj}	       
\bibliography{myref_bibtex20}

\end{document}